\documentclass[review]{elsarticle}

\usepackage[T1]{fontenc}
\usepackage[utf8]{inputenc}
\usepackage[english]{babel}
\usepackage[intlimits]{amsmath}
\usepackage[version=4]{mhchem}
\usepackage{lineno,mathtools,graphicx,braket,booktabs,multirow,siunitx,chngcntr,vmargin,xcolor,caption}
\usepackage{amssymb}
\usepackage{aas_macros}

\usepackage[pdftex]{hyperref}
\hypersetup{colorlinks=true,citecolor=red!90!black, linkcolor=black, urlcolor=blue}

\modulolinenumbers[5]

\journal{Astroparticle Physics}

\date{Last update: \today}

\newcommand{\be}{\begin{equation}}
\newcommand{\ee}{\end{equation}}
\renewcommand{\d}{\mathrm{d}}

\begin{document}

\begin{frontmatter}

\title{Atmospheric neutrinos and the knee of the Cosmic Ray spectrum}

\author{Carlo Mascaretti\corref{auth1}}
\author{Pasquale Blasi\corref{auth2}}
\author{Carmelo Evoli\corref{auth3}}
\address{Gran Sasso Science Institute (GSSI), Viale F. Crispi 7, 67100 L'Aquila, Italy \\ INFN, Laboratori Nazionali del Gran Sasso (LNGS), 67100 Assergi, L’Aquila, Italy}
\cortext[auth1]{carlo.mascaretti@gssi.it}

\begin{abstract}
The nature of the knee in the all-particle spectrum of cosmic rays remains subject of much investigation, especially in the aftermath of recent measurements claiming the detection of a knee-like feature in the spectrum of the light component of cosmic rays at energy $\sim 700$ TeV, at odds with the standard picture in which the knee in the all-particle spectrum is dominated by light cosmic rays. Here we investigate the implications of this and other scenarios in terms of the measured flux of atmospheric neutrinos. In particular we discuss the possibility that the spectrum of atmospheric neutrinos in the region $\gtrsim 50$ TeV may provide information about the different models for the mass composition in the knee region. We investigate the dependence of the predicted atmospheric neutrino flux on the shape of the light cosmic ray spectra and on the interaction models describing the development of showers in the atmosphere. The implications of all these factors for the identification of the onset of an astrophysical neutrino component are discussed.  
\end{abstract}


\end{frontmatter}

%\linenumbers

\section{Introduction} 
Despite decades of investigation on the nature of the most prominent feature in the cosmic ray (CR) spectrum, the knee remains poorly understood: a change of mass composition would suggest that the knee is due to the overlap of the contributions of accelerators running out of steam in a rigidity dependent manner, a picture that became solid after KASCADE data \cite{kascade}. On the other hand, recent measurements of the spectrum of individual elements by the ARGO collaboration suggest the existence of a knee in the light component of the cosmic ray spectrum (H+He) around $\sim 700$ GV \cite{bartoli_energy_2014}. This would imply that the knee (at few PeV) is shaped by intermediate mass elements (CNO), quite at odds with the common wisdom that associates the knee to a steepening of the spectrum of light elements. The implications of this finding for the understanding of the transition from Galactic to extragalactic CRs are also worth being mentioned. One should keep in mind the possibility that the knee might not be due to the maximum rigidity dependence in the accelerator, but rather to a change of regime in the propagation of cosmic rays through the Galaxy \cite{Giacintikachel}. This scenario implies that the maximum energy in the accelerators should be much higher than PeV, thereby increasing the tension with theoretical models of CR acceleration that have serious problems explaining how to reach even PeV energies \cite{Blasi2013}. These different possibilities, though potentially equivalent from the point of view of the all-particle cosmic ray spectrum that they result in, may lead to different predictions in terms of the spectrum of atmospheric neutrinos, which is most sensitive to the maximum energy of the light component in cosmic rays. 

In a rather qualitative way one can see that if the spectrum of Galactic CR protons were sharply cut off at energy $E_\text{max}$, the corresponding neutrino spectrum of atmospheric origin would fall at energy $\sim \xi E_\text{max}$, where $\xi\sim 0.05$. Due to either the rigidity dependence of acceleration or Galactic transport, the spectrum of He would then steepen at $2 E_\text{max}$, but the corresponding neutrino spectrum would be terminated at energy $\sim \xi E_\text{max}/2$, as determined by the energy per nucleon. The same conclusion applies to all nuclei heavier than helium (assuming a mean $A/Z\sim 2$). This simple exercise shows that the end of the spectrum of atmospheric neutrinos is mainly shaped by the parent protons rather than by heavier elements. Moreover, the flux of such heavier elements is numerically smaller, thereby making the argument above more solid. Hence, the two scenarios in which $E_\text{max}\sim \rm few$ PeV (standard model for the knee) and the one with a maximum rigidity $R_\text{max}\sim 700$ GV may result in substantially different spectra of atmospheric neutrinos in the energy region $\gtrsim 50$ TeV. 

In addition to this basic picture, one should keep in mind other aspects of the problem: 1) the knees in the spectra of individual species are not necessarily exponential, in that particle acceleration in some types of supernova remnants could lead to a spectrum of CRs with a steepening at $E_\text{max}$ rather than an exponential drop \cite{cardillo,SchureBell}; 2) the knee in the all-particle spectrum of CRs, as mentioned above, might be due to a change in the propagation regime, rather than to a rigidity dependence of the spectra of species with different mass \cite{Giacintikachel}. All these aspects are expected to affect, to different extents, the shape of the spectrum of atmospheric neutrinos in the high energy region. 

Here we calculate the atmospheric neutrino flux that follows from different assumptions on the mass composition and models of the knee and discuss the possibility to discriminate among such models by accurate measurements of the atmospheric neutrino flux, taking into account the uncertainties deriving from interaction models of CRs in the atmosphere. We also discuss the implications of a low $R_\text{max}$ (light component of CRs) in terms of the onset of astrophysical neutrinos. 

The paper is organized as follows: in \S\ref{sec:cascades} we outline the formalism relevant for the cascading of particles in the atmosphere and the ``Matrix Cascade Equation'' (MCEq) \cite{MCEQ} code that we use in the present work. In \S\ref{section:flux} we present the results obtained by modelling the knee as a cutoff and as a change of slope of the galactic CR component, and different assumptions on $R_\text{max}$. In \S\ref{sec:results} we discuss the role of uncertainties in modelling CR interactions in the atmosphere. The predictions are then compared with the IceCube measured neutrino flux \cite{ic79} and angular distribution of the events \cite{zenithdep}. The impact of the uncertainty in $R_\text{max}$ upon the onset of the astrophysical neutrino flux is discussed in \S\ref{sec:astro}. A summary of our findings and our conclusions are presented in \S\ref{sec:conclusions}.

%---------------------------------------------------------------------
\section{Equations for cascades in the atmosphere} \label{sec:cascades}

The interaction of CRs entering the Earth atmosphere induce the generation of hadronic showers, accompanied by the production of neutrinos. Following Ref.~\cite{2016crpp.book.....G}, the development of the cascade of particles of type $i$ in the atmosphere is described by solving the transport equation:
\be \frac{\d N_i(E_i,X)}{\d X}=-\frac{N_i(E_i,X)}{\lambda_i}- \frac{N_i(E_i,X)}{d_i}+\sum_{j} \int_E^\infty \d E_j\,\frac{F_{ji}(E_i,E_j)}{E_i}\frac{N_j(E_j,X)}{\lambda_j} \label{eq:casc}\ee
where $N_i(E_i,X)\d E_i$ is the flux of particles $i$ at slant depth $X$ in the atmosphere, with energy in the interval $[E_i,E_i+\d E_i]$.

The index $j$ labels $i$'s parent particles, $\lambda_i$ is the interaction length of particles of type $i$ in the atmosphere, and $d_i$ is the decay length, namely:
\[ \lambda_i = \frac{A m_p}{\sigma_{i+\text{air}}} \qquad d_i = \rho\gamma c\tau_i , \]
so that they both are in units of \si{\gram\per\square\centi\meter}; $A$ is the mean mass of target nuclei in the atmosphere and $\rho$ is the air density, which is a function of the altitude $h$. The grammage corresponding to the height $h$ can be written as

\be X(h) = \int_0^h \d\ell\,\rho(h(\ell))\label{eq:grammage} \qquad \ell = \text{particle trajectory}.\ee

The functions $F_{ji}$ are the dimensionless particle yields for the production of a particle $i$ with energy $E_i$ in a collision of a particle $j$ with energy $E_j$ on air.
They can be defined as:
\[F_{ji}(E_i,E_j) = E_i \,\frac{1}{\sigma_{j+\text{air}}} \frac{\d\sigma_{j+\text{air}\to i }}{\d E_i},\]
where $E_i$ and $E_j$ are defined in the laboratory frame.
This term can be suitably written to include the production of particles of type $i$ from the decay of their parent particle of type $j$.

While this set of coupled equations can be solved analytically in some simplified cases, such an approach would force us to assume that the spectrum of CR species is a power law (as in \cite{2016crpp.book.....G}) and would not be suitable for the investigation of the dependence of the atmospheric neutrino flux on the interaction models. 
In order to consider realistic spectra of parent CR nuclei (with a cutoff or a break at high rigidity) and to study the dependence of the results upon the physics of particle interactions in the atmosphere, we used ``Matrix Cascade Equations'' (MCEq) \cite{MCEQ} to compute the atmospheric neutrino fluxes at Earth.
MCEq is a publicly available package which allows us to adopt different CR primary spectra as well as different interaction models and compute particle cascading.

This code also features tabulated atmospheric data (e.g.~from satellites) and numerical codes, such as NRLMSISE-00 \cite{nrlm}, which we used in order to account for seasonal atmospheric variations and to average the neutrino fluxes over the zenith angle.

In the following we adopt SIBYLL-2.3c \cite{syb2.3} as our benchmark model to describe interactions in the atmosphere, while in §\ref{subsec:5} we discuss the dependence of our results on the interaction model. 

%---------------------------------------------------------------------

\section{The flux of primary CRs and the rigidity of the knee}
\label{section:flux}

The main difficulty in making physical predictions concerning the knee is that while there are very good direct measurements of individual CR species at energies $\sim 10$ TeV, our data in the energy range around the knee are rather poor. In fact, different measurements suggest rather different scenarios: the CR spectrum based upon reconstruction of KASCADE data \cite{kascade} hinted at a knee dominated by the light component (protons and He nuclei), though with a rather strong dependence of this conclusion upon the choice of the model for the description of CR interactions in the atmosphere. 
A tentative confirmation of this picture comes from the KASCADE-Grande detection of a knee feature in the heavy CR component \cite{KG} at energy $\sim 26$ times higher than the  knee in the light component.
Recent measurements by the ARGO experiment show a flux suppression (a knee) in the spectrum of light CR nuclei at energies $\sim 700$ TeV \cite{argo}, well below the energy of the knee in the all-particle spectrum. The Tibet III Collaboration also reported on the detection of a light-component knee at about \SI{500}{TeV}\footnote{See the talk by J.~Huang on behalf of the Tibet AS$\gamma$ collaboration given at ISVHECRI2018, Nagoya, Japan.}. This latter conclusion appears to confirm a tentative trend, based on older experiments as well, to locate the proton knee at somewhat lower energies for experiments at altitude closer to the maximum of the shower induced by CRs in the atmosphere (see Ref. \cite{2018book} for a recent review). This is also expected to make the dependence of the results on the adopted hadronic model weaker than for experiments at sea level. However, the interpretation of these results should take into account that, due to the very different systematics of the various experiments, their comparison is non-trivial.
 
In order to check the implications of these different scenarios for the knee, we first need to make some assumptions on the shape of the elemental spectra below the knee. The recent measurements by PAMELA and AMS-02 showed that both the spectra of protons and helium manifest a change of slope (spectral break) at the same rigidity, about 300 GV~\cite{2011Sci...332...69A,ams02i,ams02ii}. Hence we first fit
the slope and normalization of the proton and helium spectra to the AMS-02 fluxes\footnote{Publicly made available on the website \url{https://lpsc.in2p3.fr/cosmic-rays-db/}~\cite{maurin}}  above the spectral break, at $\sim 300$ GV, sufficiently far from the knee region that the assumption of power law may be considered reliable. 

In terms of the total energy $E$, we adopt the following parametrization for the power law portion of the spectrum:
\be \frac{\d N_i}{\d E} = a_i \left(\frac{E}{\SI{10}{TeV}}\right)^{-\gamma_i}\times \SI{e-7}{\per\giga\electronvolt\per\square\meter\per\second\per\steradian} \qquad i=p,\text{ He}\label{eq:pl_primary}\ee
The results of the fit for protons and helium nuclei are shown in table \ref{tab:ac}.

\begin{table}[htp]
 \centering
\begin{tabular}{ccccc}
\midrule
&&protons &&Helium \\
\midrule 
$a_i$ &&$1.5\pm0.2$ &&$1.5\pm 0.1$ \\
\midrule
$\gamma_i$ &&$2.71\pm 0.04$ &&$2.64\pm0.03$ \\
\midrule
\end{tabular}
\caption{Normalization and spectral index for protons and helium, as in Eq.~\eqref{eq:pl_primary}, as to fit AMS-02 fluxes \cite{ams02i,ams02ii}.
The reduced $\chi^2$ is $0.1$ and $0.2$ for the fit to protons ($N_\text{dof}=4$) and Helium ($N_\text{dof}=5$) respectively.}
\label{tab:ac}
\end{table}

The shape of the knee in the individual components is harder to model due to the lack of detailed measurements. Earlier measurements of the proton and helium spectra carried out by KASCADE \cite{kascade} were inconclusive in terms of locating the energy of the proton and helium knees, due to the strong dependence of the results upon the model of CR interactions in the atmosphere. It is probably more reliable to use the total spectrum of the light component as a constraint on the location and shape of the knee in the individual elements. Hence we tried to model the spectrum of light CRs at the Earth as power laws (with slope and normalization taken from fitting AMS-02 data) and a knee (modelled in two different ways) fitted to the recent ARGO and KASCADE-Grande data respectively. An exponential suppression at the knee does not provide a good fit to either set of data because of the spectral sharpness observed at the knee, hence we explored two other possibilities, namely that of an ``exponential-square'' cutoff:
\be \frac{\d N_i}{\d E}= a_i \left(\frac{E}{\SI{10}{TeV}}\right)^{-\gamma_i} \exp\left[-\left(\frac{E}{Z_i\overline R}\right)^2\right]\times \SI{e-7}{\per\giga\electronvolt\per\square\meter\per\second\per\steradian}\label{eq:exp_cut}\ee
and that of a change of slope:
\be \frac{\d N_i}{\d E}= \begin{dcases} a_{i} \left(\frac{E}{\SI{10}{TeV}}\right)^{-\gamma_{i}} \times \SI{e-7}{\per\giga\electronvolt\per\square\meter\per\second\per\steradian} &E \leq Z_i \overline R \\
b_{i} \left(\frac{E}{\SI{10}{TeV}}\right)^{-\gamma_{i}+\delta-2} \times \SI{e-7}{\per\giga\electronvolt\per\square\meter\per\second\per\steradian} &E > Z_i \overline R \end{dcases}\label{eq:spectrum_dslope}\ee
where the parameters of the fit are found using the data of KASCADE-Grande \cite{KG} and ARGO-YBJ \cite{argo} respectively. The main physical motivation for the change of slope written in the form above is that at some rigidity $\bar R$ it is expected that the diffusion coefficient changes its dependence on energy from $D(R)\propto R^{\delta}$ to $D(R)\propto R^{2}$. The transition occurs when the CR Larmor radius equals the largest scale in the turbulence $L$ that is responsible for particle scattering \cite{subedi}. For magnetic fields in the galaxy of order $\sim\mu G$ and $L\sim 10$ pc, this reflects in $\overline R\sim 3$ PeV. Smaller values of $L$ lead to correspondingly smaller values of $\overline R$. This change of slope in $D(R)$ reflects in a steepening in the CR spectrum from a slope $\gamma_{i}=\alpha+\delta$ ($\alpha$ is the slope of the injection spectrum) to $\alpha+2=\gamma_{i}-\delta+2$. This possibility has recently been discussed in \cite{Giacintikachel}. However it is worth keeping in mind that a similar change of slope might be associated to the spectrum of CRs injected by individual supernova remnants, as discussed in \cite{cardillo,SchureBell}. We notice that in Eq.~\eqref{eq:spectrum_dslope}, as in Eq.~\eqref{eq:exp_cut}, the only free parameter is $\overline R$, since $b_i$ is fixed by requiring continuity at $R=\overline R$. For the case of a change of slope, we adopt $\delta=1/3$, but we checked that a good fit can also be obtained for $\delta=1/2$. 
 
In our model for primary CRs we also included an extragalactic light component, as was measured by KASCADE-Grande \cite{KG}, with fixed slope $\gamma_\text{eg} = 2.7$ and normalization to be determined by fitting the data of KASCADE-Grande:
\be \frac{\d N_\text{eg}}{\d E }= a_\text{eg}\left(\frac{E}{\SI{100}{PeV}} \right)^{-2.7} \times \SI{e-19}{\per\giga\electronvolt\per\square\meter\per\second\per\steradian}. \label{eq:eg_comp} \ee
From the fit to the KASCADE-Grande data we obtain:
\be
a_\text{eg} = \begin{dcases} 6.0 \pm 0.2 &\text{using \eqref{eq:exp_cut}} \\ 5.0 \pm 0.5 &\text{using \eqref{eq:spectrum_dslope}} \end{dcases} \qquad \overline R = \begin{dcases} 15.1 \pm \SI{0.7}{PV} &\text{using \eqref{eq:exp_cut}} \\ 5.8 \pm \SI{0.6}{PV} &\text{using \eqref{eq:spectrum_dslope}}. \end{dcases} \label{eq:kg_fit}
\ee
The fit to the light-component data of ARGO-YBJ provides the following results:
\be \overline R = \begin{dcases} 1.3\pm \SI{0.1}{PV} &\text{using \eqref{eq:exp_cut}} \\ 640\pm \SI{50}{TV} &\text{using \eqref{eq:spectrum_dslope}}. \end{dcases}
\label{eq:argo_fit}
\ee

\begin{figure}[t]
\begin{center}
\centering
\includegraphics[width=0.49\textwidth,keepaspectratio]{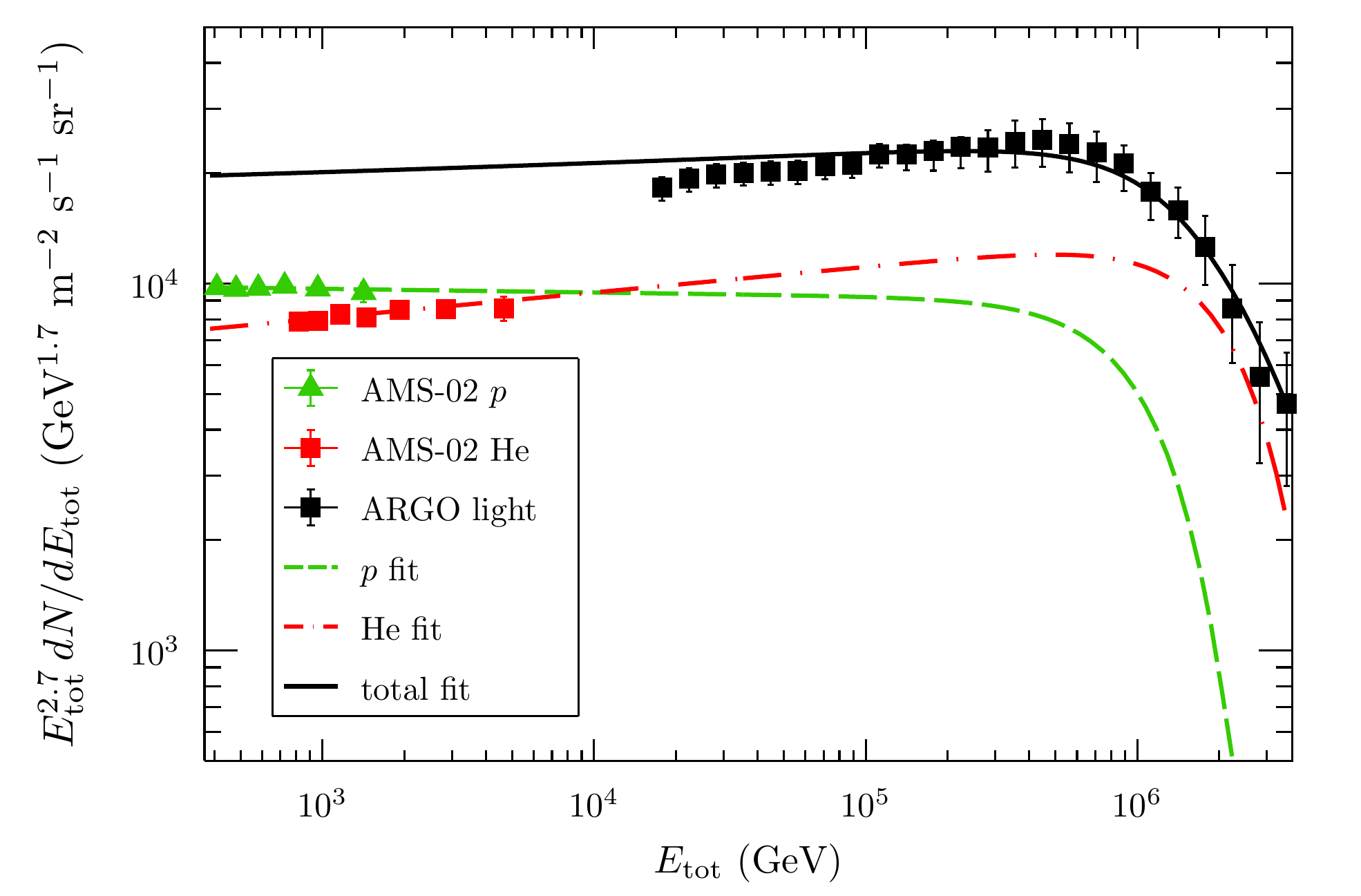}
\includegraphics[width=0.49\textwidth,keepaspectratio]{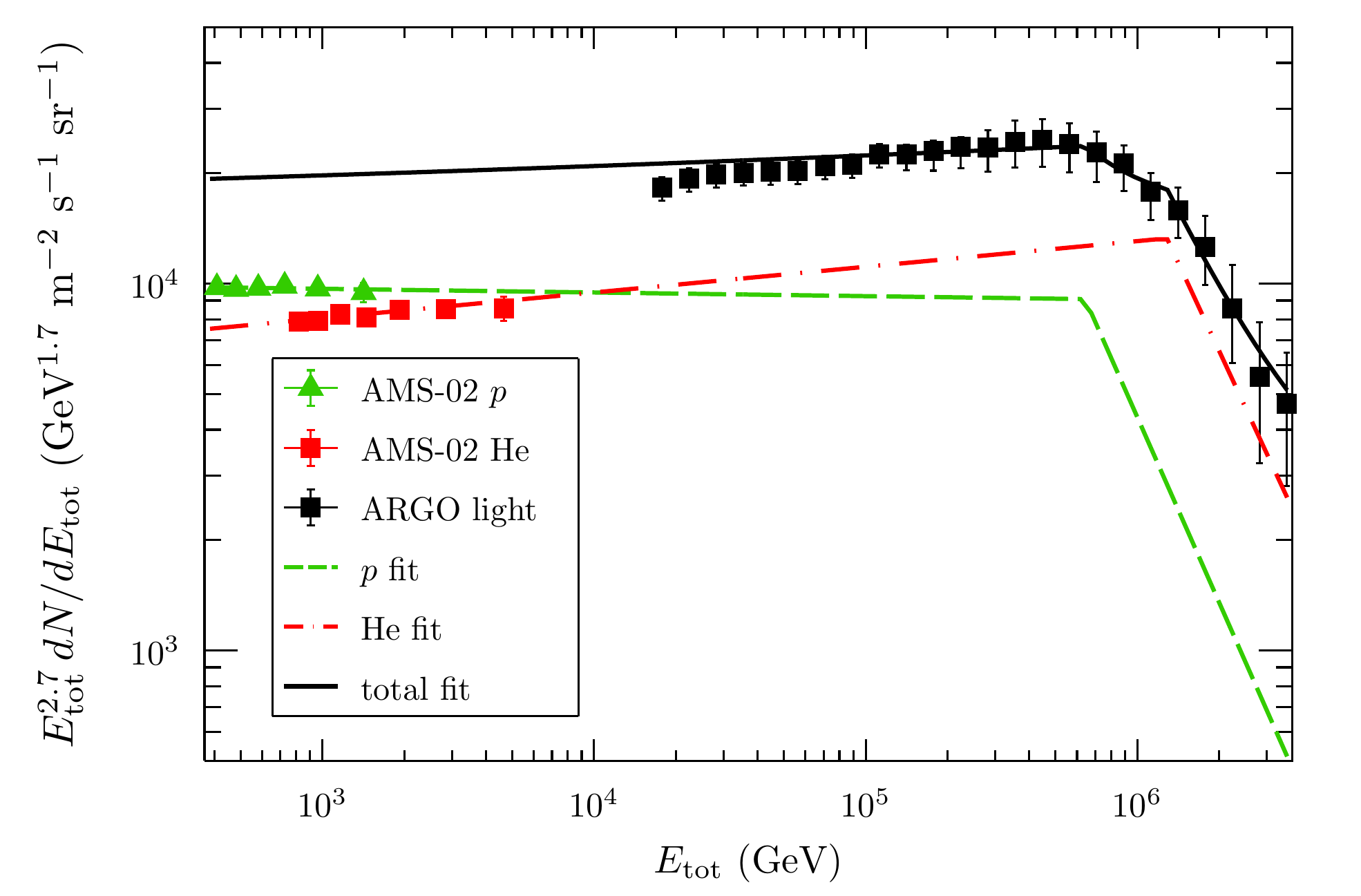} 
\end{center}
\caption{Fits to the ARGO-YBJ \cite{argo} data points with the knee modelled according to Eq.~\eqref{eq:exp_cut} (left panel) and Eq.~\eqref{eq:spectrum_dslope} (right panel). 
The low-energy part ($E \lesssim \SI{10}{TeV}$) derives from the independent power-law fit to the AMS-02 data on the spectrum of protons and helium separately, also shown in the plot.}
\label{fig:fit_argo}
\end{figure}

\begin{figure}[t]
\begin{center}
\includegraphics[width=0.49\textwidth,keepaspectratio]{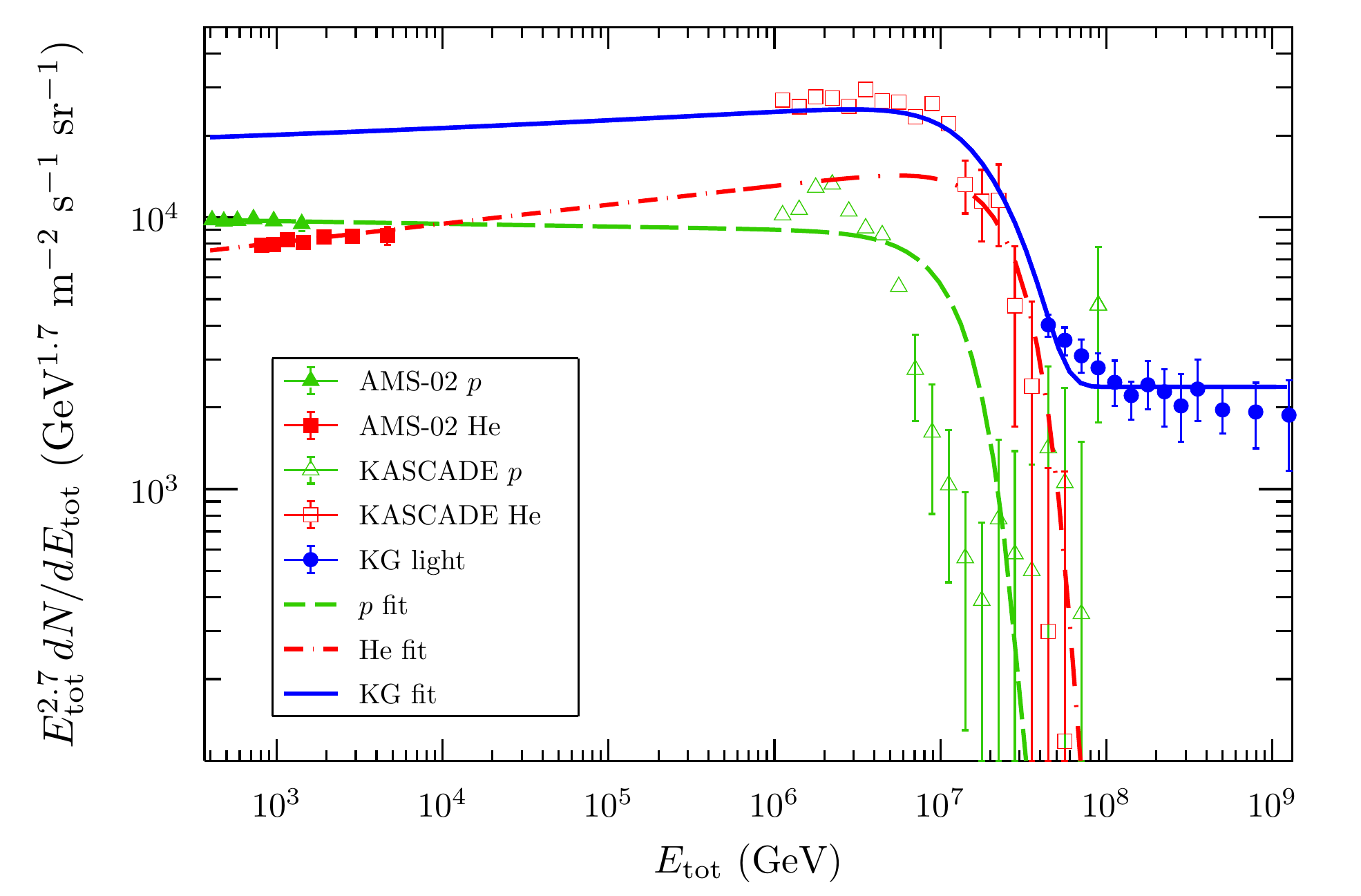} 
\includegraphics[width=0.49\textwidth,keepaspectratio]{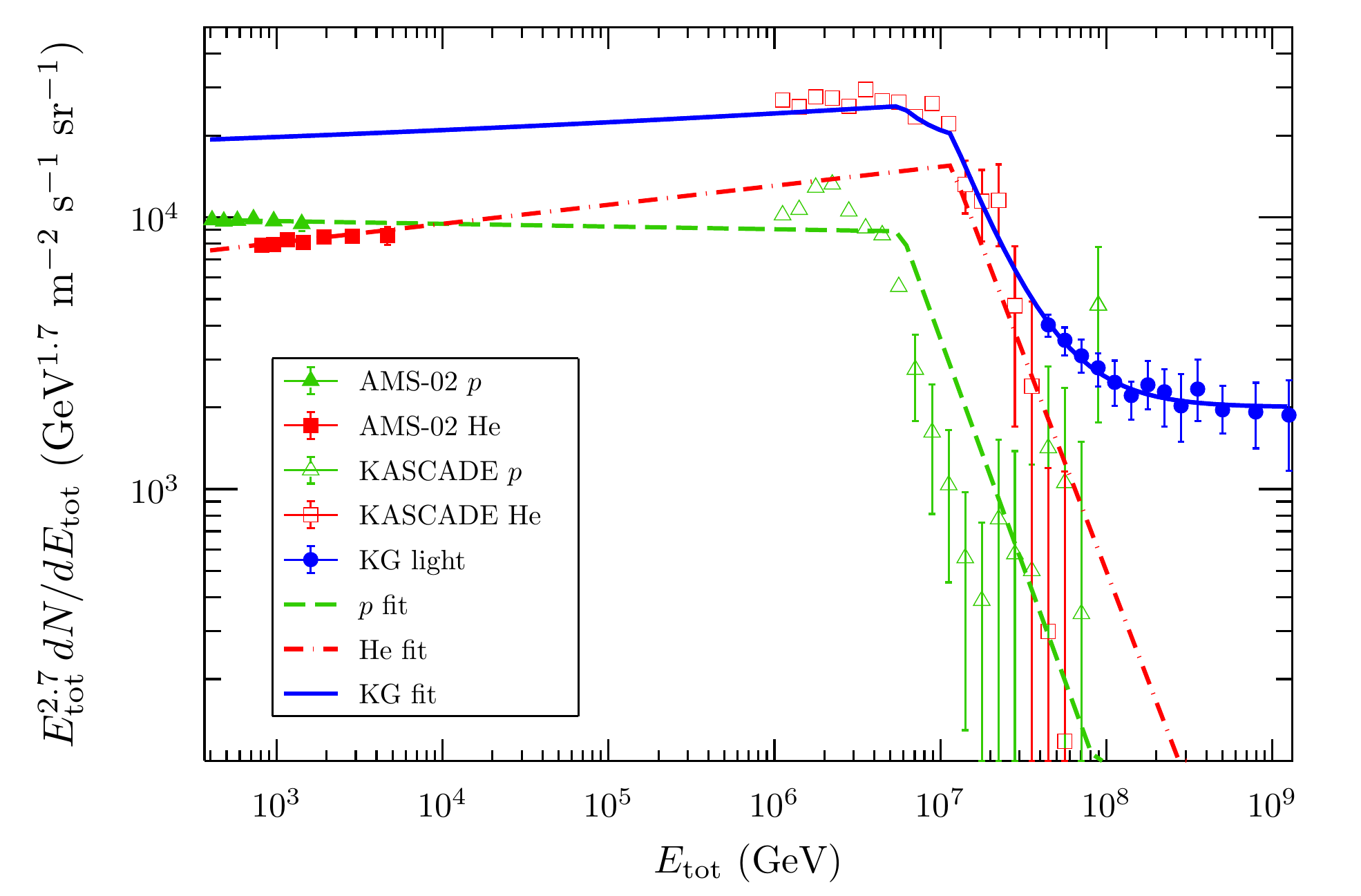} 
\end{center}
\caption{Fits to the KASCADE-Grande \cite{KG} datapoints with the knee modelled according to Eqs.~\eqref{eq:exp_cut} (left panel) and \eqref{eq:spectrum_dslope} (right panel). 
The low-energy part ($E < \SI{10}{TeV}$) derives from the independent power-law fit to the AMS-02 data.
We also show the fits to the proton, helium and total (proton $+$ helium $+$ extragalactic) spectra. 
Also shown are the fluxes of protons and helium measured by KASCADE \cite{kascade}, obtained using QGSJET as a model for CR interactions in the atmosphere.}
\label{fig:fit1}
\end{figure}

The results of the fits can be seen in Fig.~\ref{fig:fit_argo} superimposed on ARGO data and in Fig.~\ref{fig:fit1} for the KASCADE-Grande data. In both figures we also show the AMS-02 data points and the proton, helium and total (i.e.~proton $+$ helium $+$ extragalactic) fluxes. The two models of knee in the individual elements provide a reasonable fit to the data, with a slight preference for the model involving a change of slope\footnote{The change of slope fit results in a reduced $\chi^2$ of $0.1$ and $0.7$ for KG $(N_\text{dof}=11)$ and ARGO ($N_\text{dof}=23$) respectively, while the exponential-square fit results in a reduced $\chi^2$ of $0.8$ and $1.1$.} when KASCADE-Grande data are used. In Fig.~\ref{fig:fit1} we also show the proton and helium fluxes as originally derived by KASCADE \cite{kascade} using QGSJET as a model for CR interactions in the atmosphere. It is clear that data reconstructed with QGSJET are inconsistent with power-law extrapolations from the current measurements of the proton and helium spectra as measured by AMS-02 at lower energies. The same consideration applies to reconstruction with SIBYLL. It is important to realize that these codes for CR interactions in the atmosphere are not up-to-date and it would be interesting to see the KASCADE reconstructed spectra if modern versions of these interaction codes were used for the reconstruction. Some preliminary work in this direction was presented in Ref. \cite{Schoo}. Based on these considerations, we stand by our decision to fit the shape of the knees to the recent measurements of the light component (p+He) as carried out with ARGO and KASCADE-Grande respectively. 

%---------------------------------------------------------------------

\section{The atmospheric muon neutrino flux} \label{sec:results}

The flux of neutrinos of atmospheric origin is sensitive to the spectrum of parent cosmic rays. In this section we test the possibility to discriminate among different scenarios for the origin of the knee by using neutrino data. 

\subsection{Dependence on the spectrum parametrization}
\label{subsec:1}

After fitting the primary cosmic ray flux to the data, as described in the previous section, we computed the corresponding atmospheric muon neutrino flux expected in IceCube.

Clearly all mass components of cosmic rays contribute to the neutrino flux. However, protons and helium fluxes dominate the production of neutrinos. In order to demonstrate this point we compare the flux of neutrinos obtained with the protons and helium as parametrized above with the one obtained using the so-called ``H3a'' model of Hillas and Gaisser \cite{gaisser_cosmic_2013}, which comprises three different populations of five groups of nuclei each, namely:
\be \frac{\d N_i}{\d E}= \sum_{j=1}^3 \mathcal N_{i,j} E^{-\gamma_{i,j}}\exp\left(-\frac{E}{Z_i\overline R_j} \right) \qquad i=p, \text{He}, \text{CNO}, \text{Mg-Si},\text{Fe},
\label{eq:h3a} \ee
where the values of the free parameters are taken from Table 1 of Ref. \cite{h3a}. In the ``H3a'' model the knee of the light component is in the PeV region, as suggested by KASCADE observations, hence one can expect that the flux of atmospheric neutrinos in this model is closer to what we calculate for the case in which the flux of light elements is fitted to KASCADE-Grande data.  

We computed the total atmospheric muon neutrino flux at the IceCube observatory height using the MCEq code, averaging uniformly over $\cos\theta$ ($\theta \equiv$ zenith angle) and over the conditions of the atmosphere as described by the \texttt{MSIS00\_IC} model \cite{nrlm} for the South Pole in January and July. We adopted SIBYLL-2.3c \cite{syb2.3} as hadronic interaction model, unless otherwise indicated. 

\begin{figure}[t]
\begin{center}
\includegraphics[width=0.9\textwidth,keepaspectratio]{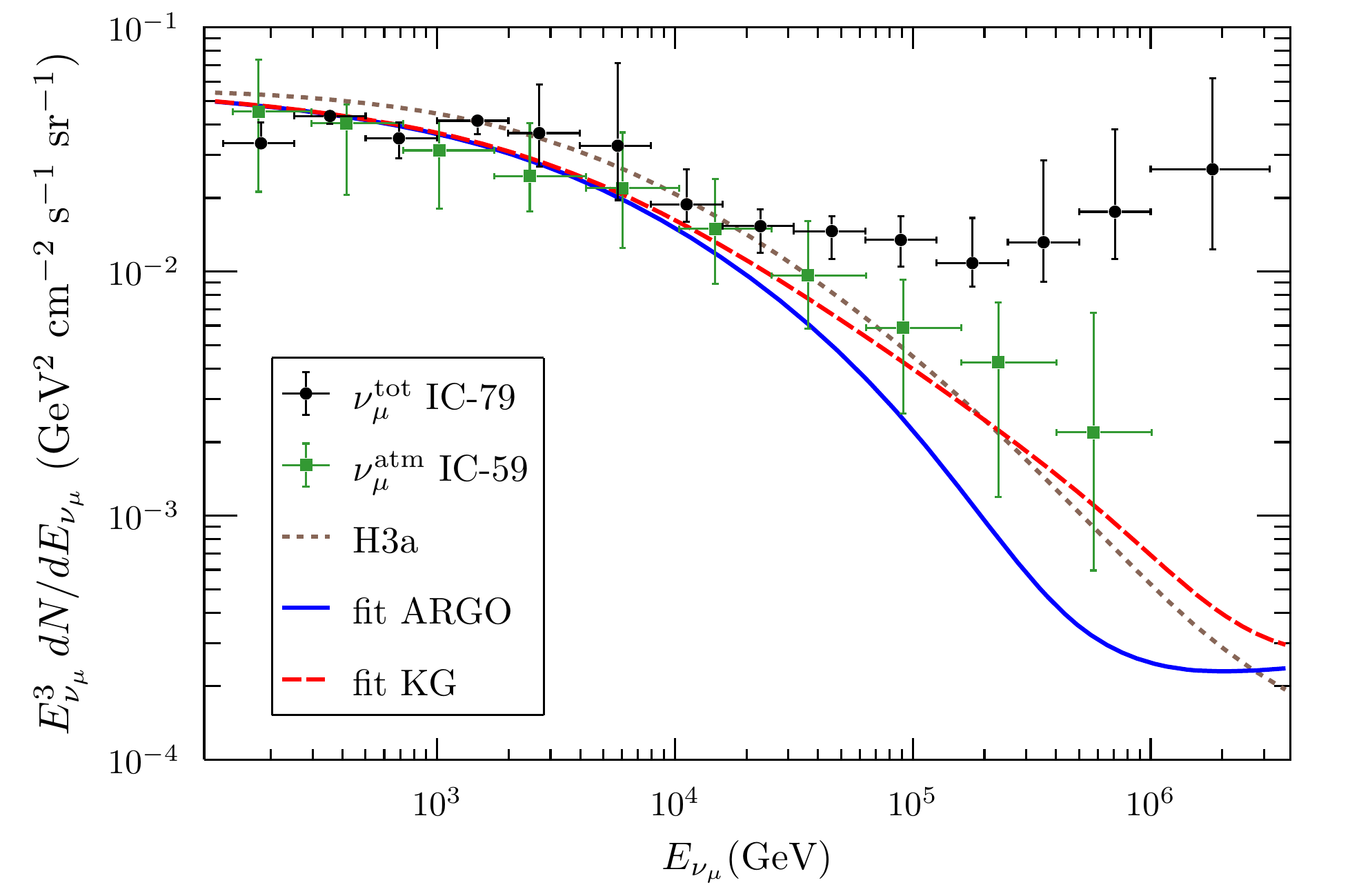}  
\end{center}
\caption{Muon neutrino fluxes resulting from the parametrizations of the primary spectrum of Eq. \eqref{eq:exp_cut}.
Our spectra are compared to those resulting from the ``H3a'' \cite{h3a} primary flux model and to the IceCube unfolded atmospheric $\nu_\mu$ spectrum \cite{ic59} and the total $\nu_\mu$ spectrum \cite{ic79}.
The vertical error bars are the the quadratic sum of the statistical and systematic uncertainties.}
\label{fig:numu_cutoff}
\end{figure}

In Fig.~\ref{fig:numu_cutoff} we show the atmospheric muon neutrino flux resulting from the primary spectrum as in Eq.~\eqref{eq:exp_cut} and for the values of $\overline R$ obtained from fitting the data of KASCADE-Grande (dashed red line) and ARGO (solid blue line). We compared the resulting fluxes to that obtained with the use of the H3a primary spectra (dotted brown line). Fig.~\ref{fig:numu_cutoff} confirms that the flux of atmospheric neutrinos is dominated by the light CR component (p+He) for a given value of the knee rigidity. In Fig.~\ref{fig:numu_cutoff} we also show the neutrino spectra as obtained by IceCube-59 \cite{ic59} and IceCube-79 \cite{ic79}. Given the smaller statistics of events, the former data points are expected to trace only the atmospheric contribution to the total neutrino flux in the high energy regime. The comparison between the solid and dashed lines show that the expected atmospheric neutrino flux is sensitive to the rigidity of the knee in the individual elements for energies above 50 TeV, while at lower energies all predictions provide an equally satisfactory description of the data. 
Although one might be tempted to express a slight preference for the primary model inspired to KASCADE-Grande data, rather than an ARGO-like model in which the knee rigidity is lower, current experimental uncertainties do not allow to draw firm conclusions.
%Although the statistical error bars are rather large, one might be tempted to express a slight preference for the primary model inspired to KASCADE-Grande data, rather than an ARGO-like model in which the knee rigidity is lower, but such considerations can only be considered tentative at the present time. 

%---------------------------------------------------------------------
\subsection{Theoretical uncertainties}
\label{subsec:5}

There are two types of theoretical uncertainties that affect the calculation of the muon neutrino flux, namely uncertainties in the parameters describing the flux of primary CRs (see table \ref{tab:ac}) and uncertainties deriving from the choice of the hadronic interaction model.

\begin{figure}[t]
\begin{center}
\includegraphics[width=0.9\textwidth,keepaspectratio]{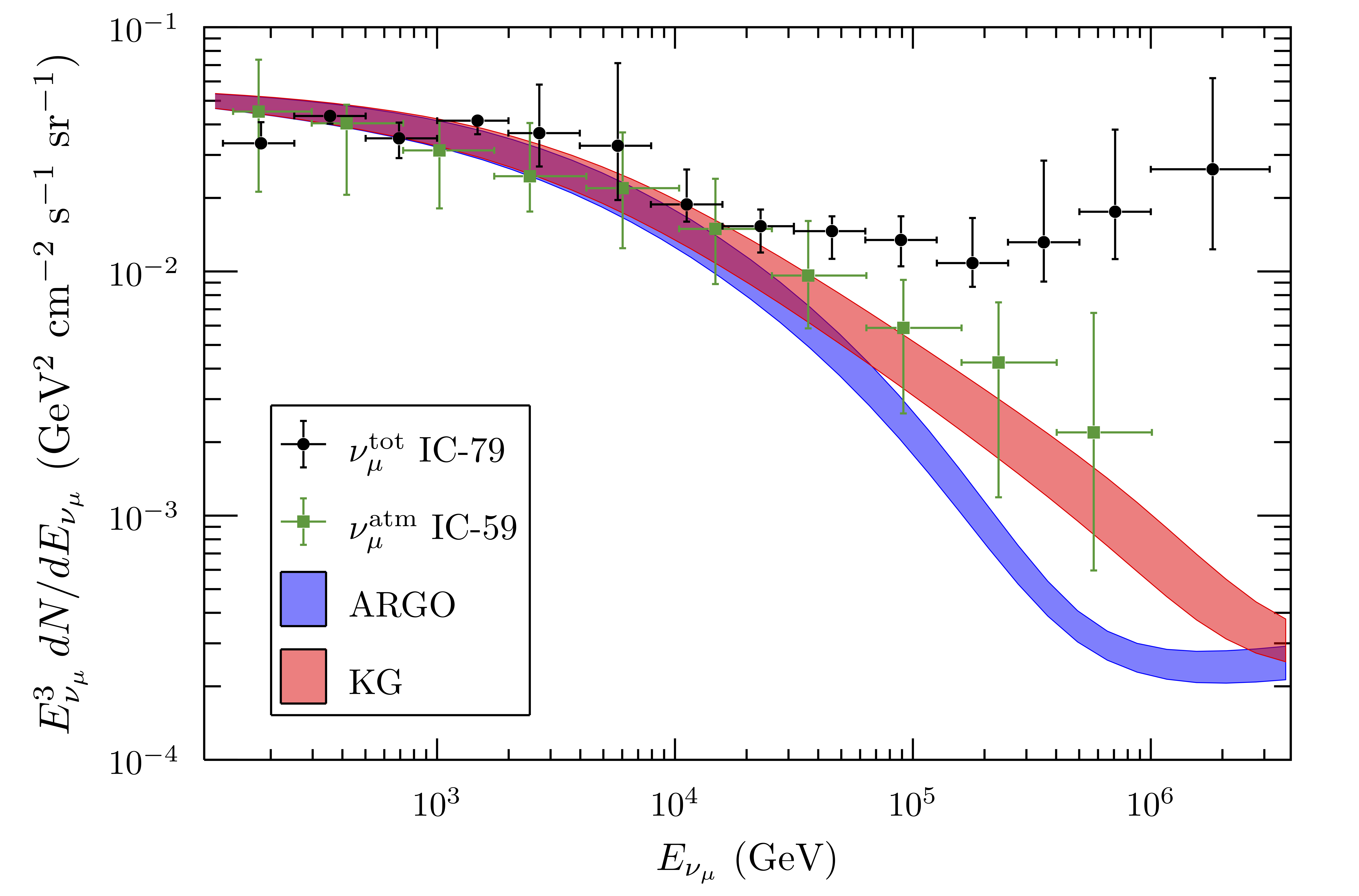} 
\end{center}
\caption{The atmospheric muon neutrino flux uncertainty due to that on the primary cosmic ray flux parameters and on its functional form: the bands are delimited by the largest  and the smallest fluxes obtained by choosing $(a_i+\delta a_i;\gamma_i-\delta \gamma_i;\text{exp-square-knee})$ and $(a_i-\delta a_i;\gamma_i+\delta \gamma_i;\Delta\gamma-\text{knee})$ respectively.
These fluxes are compared to the IceCube unfolded $\nu_\mu$ spectrum \cite{ic79} and to the unfolded atmospheric $\nu_\mu$ spectrum \cite{ic59}.}
\label{fig:bands}
\end{figure}

The former can be quantified by calculating the minimum and maximum flux of atmospheric neutrinos obtained by changing the parameters describing the fluxes of primary protons and helium (with different assumptions on the shape of the knees). The result of this calculation is illustrated in Fig.~\ref{fig:bands} for the case of ARGO-like and KASCADE-Grande-like knee. The shaded bands illustrate the uncertainties deriving from the shape of the spectrum at low energy and the shape of the knees. Given such uncertainties, it appears that a separation between the low rigidity and high rigidity cases is possible for neutrino energies above $\sim 100$ TeV, although in that energy region the current statistics of events is rather low and the contribution of astrophysical neutrinos to the total flux is important. 
With all these caveats, we computed the average residual of the IC-59 \cite{ic59} data with respect to the top of the KG and ARGO band for the 5 most energetic datapoints: we obtain $0.9$ for KG and $1.5$ for ARGO, which shows some weak preference for the case with high rigidity knee in the light CR component. 

\begin{figure}[t]
\begin{center}
\includegraphics[width=0.9\textwidth,keepaspectratio]{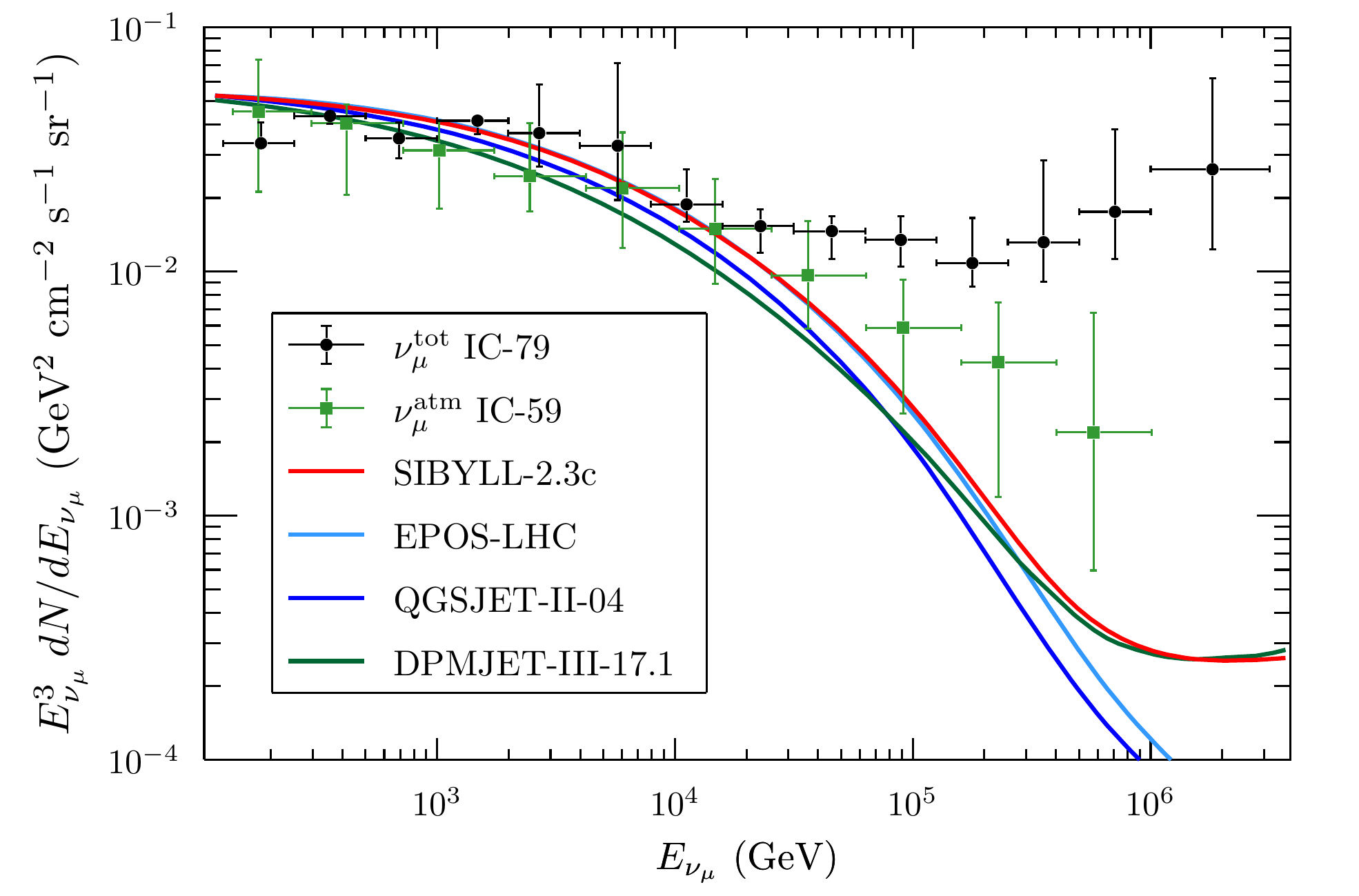} 
\end{center}
\caption{Muon neutrino fluxes resulting from four interaction models available in MCEq \cite{syb2.3,eposlhc,qgsjet,dpmjetiii}. As primary spectrum we used Eq.~\eqref{eq:spectrum_dslope} fitted to the ARGO datapoints with normalization $a_i+\delta a_i$ and slope $\gamma-\delta\gamma_i$ in order to obtain the largest muon neutrino flux possible according to the ARGO data. These fluxes are compared to the IceCube unfolded $\nu_\mu$ spectrum \cite{ic79} and to the unfolded atmospheric $\nu_\mu$ spectrum \cite{ic59}.}
\label{fig:dep_had_mod}
\end{figure}

In order to quantify the dependence of our results on the interaction model, we computed the muon neutrino fluxes employing four hadronic interaction models available in MCEq, namely SIBYLL-2.3c \cite{syb2.3}, EPOS-LHC \cite{eposlhc}, QGSJET-II-04 \cite{qgsjet} and DPMJET-III-17.1 \cite{dpmjetiii}. Our results are shown in Fig.~\ref{fig:dep_had_mod}, together with the IC-59 and IC-79 data points. The difference in the theoretical predictions at energies $\gtrsim 100$ TeV are due to that fact that QGSJET and EPOS do not include the contribution of prompt neutrinos. 

We assumed a primary spectrum like Eq.~\eqref{eq:spectrum_dslope}, fitted to the ARGO data, and with normalization $a_i+\delta a_i$ and slope $\gamma_i-\delta\gamma_i$ in order to maximize the atmospheric neutrino flux in the case of a fit to the ARGO data. The aim of this exercise is to check the extent to which the difference between KASCADE-Grande and ARGO fits to light primary CR can be masked by the uncertainties in the interaction models. It appears that the uncertainties due to the fit to primaries and those deriving from interaction models are comparable. 

Another source of uncertainty in the atmospheric neutrino flux is the contribution of the prompt component, namely neutrinos due to the decay of charmed mesons produced in cosmic rays collisions on the atmosphere, which is yet to be measured. Many (semi-)analytical computations \cite{TIG,sarcevic,honda,rottoli,garzelli,berss} have been carried out, adopting different primary CR spectra and hadronic interaction models. Our predictions based on using MCEq, adopting the primary CR fluxes as defined in Sec.~\ref{section:flux} and adopting SYBILL-2.3c as interaction model, agree with the most recent of these computations. As can be seen from Fig.~\ref{fig:dep_had_mod}, the level of uncertainty due to the prompt component becomes somewhat of a concern at $\gtrsim\SI{300}{TeV}$, so that it is not expected to affect in any significant way our conclusions on the position of the knee in the light component.

%---------------------------------------------------------------------
\subsection{Angular distribution expectations}
\label{subsec:6}
A safe discrimination between different models of the knee in the individual light elements requires neutrinos with energies above a few hundred TeV and a clear tagging of atmospheric neutrinos, perhaps based upon the angular distribution of the signal. In fact neutrinos of astrophysical origin are expected to show a quasi-isotropic angular distribution. Such isotropy may either reflect the homogeneity of the universe on cosmological scales (the pathlength of neutrinos at the energies of interest is larger than the size of the universe), if the sources have a cosmological spatial distribution~(see for instance~\cite{halzen} and references therein), or the presence of a large emission region around our own Galaxy, as would be the case in some models \cite{2014PhRvD..89j3003T,BlasiAmatoPRL}. 

\begin{figure}[t]
\begin{center}
\includegraphics[width=0.9\textwidth,keepaspectratio]{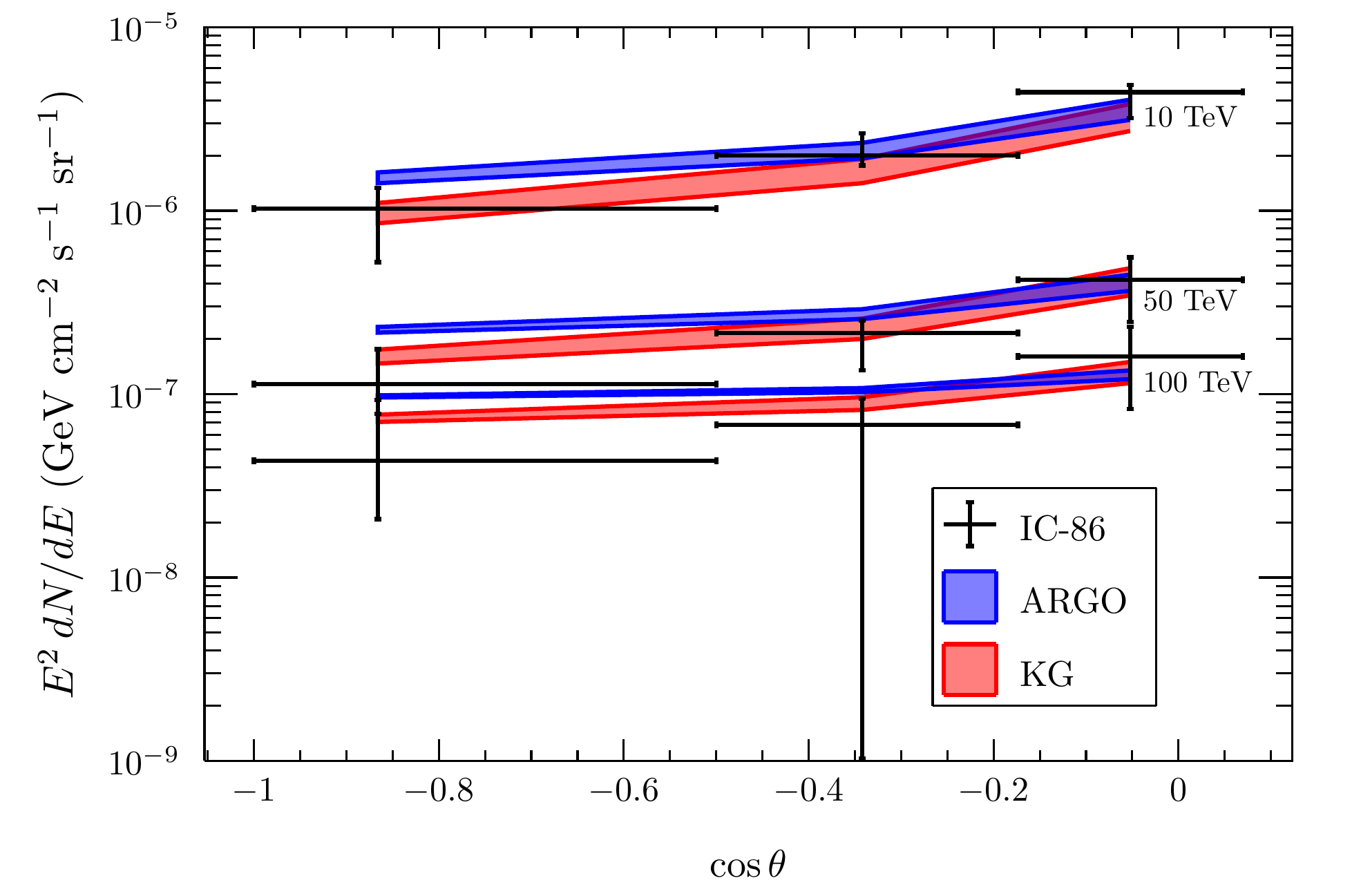} 
\end{center}
\caption{The muon neutrino fluxes as a function of the zenith angle $\theta$ for $E_\nu = 10, 50, 100$ TeV resulting from the primary models fitted to the KASCADE-Grande (KG) and ARGO datapoints compared to the measured flux of \cite{zenithdep} (IC-86).
We averaged our neutrino fluxes in the same angular bins for which the data are reported in \cite{zenithdep}. The theoretical uncertainties on the primary spectrum are shown (shaded areas). An astrophysical neutrino flux has been added to the atmospheric contribution in order to fit the total IC-86 preliminary flux.}
\label{fig:zenith}
\end{figure}

Some information on the observed angular distribution of neutrinos was recently presented in Ref.~\cite{zenithdep} for IC-86; unfortunately in the highest energy bin used in the analysis, around $100$ TeV, the expected effect is still rather marginal. 
This is clearly visible in Fig.~\ref{fig:zenith}, where we show the data of IC-86 \cite{zenithdep} compared with our predictions for a low and high rigidity model of the knee in the light component. Notice that here an astrophysical neutrino flux was added to the atmospheric contribution (including prompt neutrinos) so that the total flux is a good fit to the preliminary IC-86 data points. We will comment below on such an astrophysical component. Visual inspection of Fig.~\ref{fig:zenith} clearly shows that data on the angular distribution of the events do not have much discrimination power between different models of the knee, at least at the energies considered so far. 

%---------------------------------------------------------------------
\section{Astrophysical neutrinos}
\label{sec:astro}

The difference between the neutrino flux measured by IC-79 and the expected atmospheric neutrino flux (including prompt neutrinos) provides evidence for an additional component that is naturally interpreted to be of astrophysical origin. From the discussion above it is clear that the modelling of the shape and position of the knees in the light CR component affects the inferred spectrum of neutrinos of astrophysical origin, which makes the motivation for accurate understanding of atmospheric neutrinos even stronger. 

For the model in which the knee is at higher energy (motivated by KASCADE-Grande data) our best fit to the differential spectrum of the additional component can be written as
\begin{equation}
\frac{\d\Phi_{\nu}}{\d E_{\nu}} = (7\pm 3) \times 10^{-18} \left( \frac{E_{\nu}}{\SI{100}{TeV}}\right)^{-2.6\pm0.2} ~ \si{\per\giga\electronvolt\per\square\centi\meter\per\second\per\steradian},
\end{equation}
while for a lower energy knee of the light component (ARGO-like) the best fit that we obtain is
\begin{equation}
\frac{\d\Phi_{\nu}}{\d E_{\nu}} = (10\pm 3) \times 10^{-18} \left( \frac{E_{\nu}}{\SI{100}{TeV}}\right)^{-2.9\pm0.2} ~ \si{\per\giga\electronvolt\per\square\centi\meter\per\second\per\steradian}.
\end{equation}
with reduced $\chi^2$ of $0.06$ in both cases ($N_\text{dof}=3$).
It is easy to understand that in the latter case the atmospheric component is suppressed at lower energies, so that the inferred astrophysical neutrino spectrum needs to be steeper than in the former case and with a higher normalization at 100 TeV. 

The IceCube collaboration has released the results of different analyses revealing the detection of an astrophysical component of the neutrino flux. The dataset based on muon tracks has been fitted to the expression \cite{MuonTracks}:
\begin{equation}
\frac{\d\Phi_{\nu}^\text{tracks}}{\d E_{\nu}} = (0.9  ^{+0.30}_{-0.27}) \times 10^{-18} \left( \frac{E_{\nu}}{\SI{100}{TeV}}\right)^{-2.13\pm0.13} ~ \si{\per\giga\electronvolt\per\square\centi\meter\per\second\per\steradian},
\end{equation}
while high energy starting events (HESE) selected with energies $\gtrsim 60$ TeV in the last six years appear to have a substantially steeper spectrum \cite{hese}:
\begin{equation}
\frac{\d\Phi_{\nu}^\text{HESE}}{\d E_{\nu}} = (2.46\pm0.8) \times 10^{-18} \left( \frac{E_{\nu}}{\SI{100}{TeV}}\right)^{-2.92^{+0.33}_{-0.29}} ~ \si{\per\giga\electronvolt\per\square\centi\meter\per\second\per\steradian}.
\end{equation}
In both cases the adopted atmospheric neutrino flux is expected to be close to the one derived in our scenario with a higher energy knee in the light CR component. Yet, the inferred flux of astrophysical neutrinos at 100 TeV in IceCube turns out to be a factor $\sim 2-3$ below the one inferred above and based on our calculated atmospheric neutrino flux, which is in good agreement with other calculations present in the literature, as the conventional atmospheric best fit in Ref.~\cite{MuonTracks}. In part this discrepancy may be due to the anomalously large neutrino flux in IC-79 (about a factor 2 above average for a reconstructed muon energy $\gtrsim 100$ TeV, but still compatible within statistical fluctuations). 
On the other hand, it seems likely that the atmospheric neutrino flux adopted in the IceCube analyses may be somewhat different from that typically used in the literature. This type of details may be difficult to assess with a scrutiny from outside the collaboration, hence we argue that more details on this point might in fact be useful to the community to make an independent assessment of the actual flux of neutrinos having a non-atmospheric origin. 

%---------------------------------------------------------------------
\section{Conclusions} \label{sec:conclusions}

The end of the light mass components of Galactic CRs contains much information on both acceleration and transport, but the measurement of this part of the spectrum has led to results that are far from conclusive: for instance, the KASCADE measurement \cite{kascade} of the proton and helium spectrum at energies $\gtrsim$ PeV led to the detection of a knee-like feature at rigidity of a few PV, consistent with being the result of the maximum energy in the accelerators. However, the existence of such a feature and the energy at which it is located depend upon the model adopted to describe CR interactions in the atmosphere. All models used at that time are now considered obsolete. More recently the ARGO experiment measured the spectrum of the light CR component and observed a knee-like feature at energies $\sim 700$ TeV, quite at odds with the previous result of the KASCADE experiment. At the same time, KASCADE-Grande, a higher energy evolution of the old KASCADE experiment, measured the spectrum of the light and heavy CR components \cite{KG}: the decrease of the light component and the position of the knee in the heavy component are consistent with having a knee in the light CR spectrum in the PeV region. In any case, the shape of the knee (cutoff versus a change of slope) remains uncertain. 

It is worth stressing here that the position of the knee in the light CR spectrum is of the utmost importance also for the physical description of the transition from Galactic to extragalactic CRs: in the standard picture of Galactic CRs, the spectrum of the individual elements gets suppressed (in a model dependent way) at the maximum rigidity of the accelerator, so that the Galactic CR spectrum is expected to end with a heavy composition at energies $\sim 10^{17}$ eV. This picture seems supported by the KASCADE-Grande measurement of the iron spectrum \cite{KG}, which shows a knee-like feature at such energies. The issue of the transition to extragalactic CRs in the aftermath of KASCADE-Grande data was discussed in detail in Ref.~\cite{BereAloBla}. On the other hand, the issue of the transition in an ARGO-like scenario for the knee remains problematic and quite poorly discussed. New experiments at high altitude like LHAASO \cite{lhaaso} could provide us with important information about the origin of the knee.

Since the end of the spectrum of atmospheric neutrinos is supposed to retain memory of the end of the light CR spectrum (protons and helium nuclei), here we discussed the effect of a KASCADE-like and ARGO-like knee and of the shape of such knee for the spectrum of atmospheric neutrinos. We also investigated the uncertainties induced on the atmospheric neutrino flux due to the adoption of different models for the interactions of CRs in the atmosphere. We found that the uncertainties due to the shape of the CR spectrum and those due to interaction models are comparable once the position of the proton knee is fixed. Throughout most of our calculations we adopted SIBYLL-2.3c \cite{syb2.3} as hadronic interaction model. This model includes the production of prompt neutrinos. 

A clear distinction between an ARGO-like and a KASCADE-like knee seems possible at energies $\gtrsim 100$ TeV if the atmospheric neutrinos could be properly tagged. Unfortunately this is also the energy region where the total neutrino flux detected by IceCube departs from the existing predictions for atmospheric neutrinos. This is usually interpreted as the onset of a neutrino component having an astrophysical origin. So far, the sources of such neutrinos remain unknown (see \cite{halzen} for a review). The IceCube fits to the spectrum of astrophysical neutrinos vary depending upon the analysis performed. The spectral slope is mainly determined by the lower energy events of the extra component, where the atmospheric contamination is the highest. It is therefore of crucial importance to model the end of the atmospheric neutrino flux in a credible way in order to assess with confidence the astrophysical nature of the extra component and to characterize it appropriately. As we discussed above, the flux of atmospheric neutrinos in the energy region $\gtrsim 100$ TeV is affected by both an uncertainty due to the shape of the proton knee and uncertainties in the hadronic interaction model. 
An additional source of uncertainty is due to the calculation of prompt neutrinos. It remains true however that the largest limiting factor in our knowledge of the atmospheric neutrino contribution and the transition to astrophysical neutrinos is the poor understanding of the physics of the knee. In this sense a dedicated analysis of the IceCube data in the perspective of identifying genuinely atmospheric neutrinos in the high energy region ($E\gtrsim 100$ TeV) would be of the utmost importance. 

\section*{Acknowledgments}

The authors are grateful to Anatoli Fedynitch, Ivan De Mitri, Paolo Lipari, Giulia Pagliaroli, Francesco Villante, and Francesco Vissani for several discussions on the topics illustrated here, and to the anonymous referees for the useful and detailed comments. PB is grateful to P. Mertsch, C. Wiebusch and C. Haack for hospitality and useful discussion in Aachen. The research of PB was partially funded through Grant ASI-INAF n.~2017-14-H.0. CE~acknowledges the European Commission for support under the H2020-MSCA-IF-2016 action, Grant No.~751311 GRAPES Galactic cosmic RAy Propagation: an Extensive Study.

\section*{References}
\bibliographystyle{apj}
\bibliography{muons}

\end{document}